\documentclass[final,journal]{IEEEtran}
\usepackage{amsmath,amsfonts}
\usepackage{algorithmic}
\usepackage{algorithm}
\usepackage{array}
\usepackage{textcomp}
\usepackage{url}
\usepackage{verbatim}
\usepackage{graphicx}
\usepackage{subfigure}
\usepackage{cite}
\usepackage{amssymb}
  \usepackage{multirow}

\usepackage[breaklinks,backref]{hyperref}

\usepackage{doi}
\usepackage{enumerate}

\def\d{\mathrm{d}}
\def\f{\mathrm{f}}

\def\b{\mathrm{b}}
\def\e{\mathrm{e}}
\def\m{\mathrm{m}}

\def\org{\mathrm{org}}

\begin{document}

\title{Comprehensive Energy Footprint Benchmarking of Strong Parallel Electrified Powertrain}
\author{Hamza~Anwar*,~\IEEEmembership{Student Member,~IEEE}\thanks{Manuscript received \today; revised \today. This paper has not been presented at any conference, and is not under consideration for publication anywhere else. This work was performed as part of an industry sponsored collaborative research project at the Center for Automotive Research, Ohio State University.}
\thanks{*Corresponding Author: Mr. Hamza Anwar, 930 Kinnear Road, Columbus, OH 43212, email: \texttt{anwar.24@osu.edu}.
PhD Candidate in Electrical Engineering at the Ohio State University
.},
\and
Aashrith~Vishwanath%
, 
\and
Qadeer~Ahmed,~\IEEEmembership{Senior Member,~IEEE}%
, 
\and
Apurva~Chunodkar%
}

\markboth{P\MakeLowercase{reprint submitted}}{}%


\maketitle
\begin{abstract}
This work presents comprehensive energy management and in-depth energy footprint analysis of an electrified strong parallel commercial vehicle. We use the PS3 framework, validated real-world powertrain system models, and Pareto-optimal analysis to optimize fuel consumption and harmful pollutant emissions. The approach involves dynamic optimization of 13 states and 4 control levers with complex interactions between multiple subsystems for a parallel hybrid electric pick-up and delivery truck. These subsystems exhibit thermal, electrical, and mechanical dynamics at different time scales, and contain kinematic and combinatorial constraints, integer- and real-valued variables, interpolated look-up tables, and data maps. A Pareto-optimal solution is found by carefully optimizing fuel and NOx emissions to understand the energy footprint of the electrified powertrain. The presented results exhibit rich analysis and complex interactions among the powertrain subsystems to unearth a 7$\%$ improvement in its fuel consumption and 29$\%$ pollutant NOx reduction when compared to solution from a coarsely modeled powertrain system.
\end{abstract}

\IEEEpeerreviewmaketitle
\section{Introduction}
\label{sec:intro}
\IEEEPARstart{T}{he} transportation sector is responsible for more than 29\% of greenhouse gas (GHG) emissions and over 55\% of total NOx emissions in the U.S. \cite{epaCleaner2} where the largest NOx pollutant contributors are commercial medium and heavy duty trucks. In response, the developed world sees increasing numbers of battery electric vehicles on the road \cite{del2020} but when it comes to commercial trucks, the benefits of strong hybrid electrified vehicles (HEV) which combine the pros of conventional and battery electric vehicles still arguably outweigh \cite{Zhou2014}. Owing to the promising future of commercial hybrid trucks, research on energy management strategies (EMS) is a growing area \cite{tran2020thorough,sabri2016review}. With intricate interactions between the complex powertrain subsystems, diverse scope of variables, and multiple objectives, optimal energy management in HEVs for comprehensive solutions becomes involved. Popular optimization-based energy management approaches include Dynamic Programming (DP) \cite{Brahma2000,Maddumage2021}, Pontryagin's Minimum Principle \cite{Nguyen2021,1}, numerical optimization \cite{9861632,4,Wu2020}, and their hybrids \cite{3}, each with drawbacks related to curse of dimensionality, intractability of path constraints, assumption of linear or convex models, or inability of handling integer variables. There are hardly any works on \textit{simultaneously} optimizing performance of numerous powertrain subsystems (such as battery, combustion engine, traction motor, after-treatment, transmission) with larger number of associated mixed-integer and nonlinear state and control variables in an optimal control fashion. Works on complete vehicle energy management (CVEM) \cite{kessels2012smart,padilla2022complete} try to bridge this gap but are limited by their assumptions of linear state dynamics or quadratic energy inflow-outflow relationships. In literature, we have not found any optimal powertrain control work solving a problem with more than handful number of states and controls. Furthermore, when it comes to conflicting cost function terms,  works focusing on joint fuel and NOx pollutant emissions minimization \cite{8068210,Feng2022,10} are also few in number and similarly lack in system-wide comprehensiveness. 

In this paper, we present a case-study of comprehensive energy management optimization for a 13-state 4-control powertrain problem in a class-6 strong parallel P2 hybrid electric pickup and delivery truck. Complex interactions between the validated models of powertrain subsystems is included and results for three scenarios of diesel fuel and NOx emissions minimization after Pareto-front study are presented along with detailed component-wise energy analysis. The control problem consists of some fast dynamics like battery state-of-charge (SOC), some slow dynamics like battery temperature and catalyst temperatures in after-treatment system, some discrete dynamics like gear selection and engine on/off status, and some continuous dynamics like vehicle acceleration. Formulations imposing discontinuities, the use of real-world data maps of the engine, motor, battery, and after-treatment systems, and problem stiffness make our case-study problem challenging. Additional challenges considered are of complicating boundary and path constraints such as battery charge sustaining and vehicle speed modulation, i.e., eco-driving, and minimum dwell-time on engine status and gear switching. Objective function trades-off between overall fuel consumption and system-out NOx emissions. A comprehensive optimization strategy catering for all mixed-integer nonlinear considerations described above has not been applied in literature.

To solve our case-study problem, we use the novel PS3 approach which is presented in our prequel work \cite{paper1citation}.
The PS3 algorithm is a three-step \textit{first-discretize-then-optimize} direct method of numerical optimization for powertrain control problems that uses pseudo-spectral collocation for highly accurate state estimation. It breaks the mixed-integer optimal control problem into two nonlinear programming (NLP) and one mixed-integer quadratic programming (MIQP) problems. Depending on how and when they are solved in the three steps of PS3, the state and control variables are categorized into \textit{consistent} continuous, discrete, and \textit{inconsistent} continuous types | see our prequel for formal description and algorithm details. Our current work in this paper focuses on a specific case-study problem and application, not on the algorithm.
\subsection{Case-Study Overview and Paper Organization}
\label{sec:Prb_des&form}
\begin{figure}[!t]
	\centering
	\includegraphics[width=\linewidth]{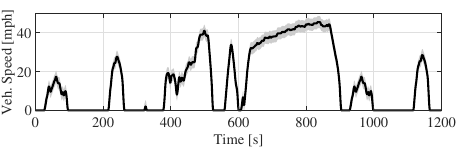}
	\includegraphics[width=\linewidth]{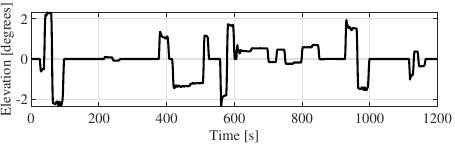}
	\caption{Reference speed profile with constraint envelope (top), elevation profile (bottom).\label{fig:vehdyn}}
\end{figure}

\begin{figure}[!t]
	\centering
	\includegraphics*[width=\linewidth]{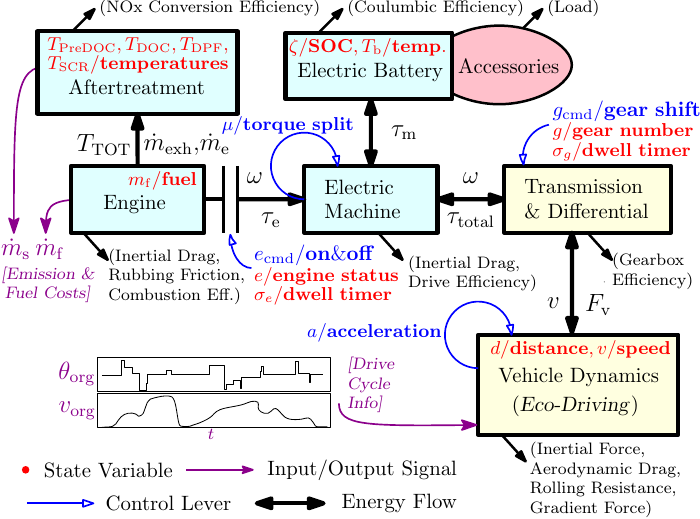}
	\caption{Case-study problem block diagram.\label{fig:blockdiag}}
\end{figure}

We consider a class-6 pick-up and delivery truck with strong P2 parallel hybrid architecture. This vehicle operates on a reference real-world urban pickup and delivery duty cycle known \textit{apriori}, shown in Fig.~\ref{fig:vehdyn}. 
The block diagram in Fig.~\ref{fig:blockdiag} depicts the case-study architecture, energy flow, various sub-components, their states, controls and other signals described later in the paper.
There are 13 states and 4 control variables. The state variables and control variables with their types and symbols are given in Table \ref{tab:S&C}. These variables are optimized across the three steps of the PS3 algorithm. The cost function for this optimization problem is divided into three cases depending on the value of the weighing factor $\beta \in [0,1]$, and is expressed as:
\begin{equation}\label{eq:cost_func}
    J:=\int_{0}^{T}\beta\dot{m}_{\mathrm{f}}+(1 - \beta)\dot{m}_{\mathrm{s}}\, \d t,
\end{equation}
where, $T=1200$ seconds is the drive cycle duration, $\dot{m}_\mathrm{f}$ the rate of fuel consumption, and $\dot{m}_{\mathrm{s}}$ the rate of system-out NOx emissions: (a) \textit{Fuel problem} minimizes only the fuel consumption, $\beta=1$, (b) \textit{Emissions problem} minimizes only system-out NOx pollutant emissions, $\beta=0$, (c) \textit{Fuel \& Emissions problem} minimizes a combination of fuel consumption and system-out NOx emissions wherein $\beta$ is chosen appropriately through Pareto-front study presented in Section \ref{subsec:pareto study}. Modeling details that explain continuous and discrete control action and dynamics of states are given in Section \ref{sec:modeling}, which also includes formal definitions of various path, box, bound, initial- and final-value constraints. Implementation details in Section \ref{sec:Implementation} explain the numerical programs that the PS3 algorithm formulates and solves. Detailed results followed by energy analysis is presented in Section \ref{sec:Results}.
\begin{table}[!t]
	\caption{State and control variables with symbols and types\label{tab:S&C}}
	\centering
	\begin{tabular}{m{0.42\linewidth}m{0.10\linewidth}m{0.33\linewidth}}
		\hline\hline
		\textbf{States (13)} & \!\!\textbf{Symbol} & \textbf{Variable Type (PS3)}\\
		\hline
		Vehicle speed &\!\!$v$ &\multirow{2}{*}{Continuous (consistent)}\\
		Vehicle distance & \!\!$d$ & \multirow{2}{*}{} \\
		\hline
		Engine status & \!\!$e$ & Discrete (binary-valued)\\
		\hline
		Gear number & \!\!$g$ & \multirow{3}{*}{Discrete (integer-valued)}\\
		Engine on/off dwell-time counter & \!\!$\sigma_e$& \multirow{3}{*}{}\\
		Gear dwell-time counter & \!\!$\sigma_g$& \multirow{3}{*}{}\\
		\hline
		Battery state-of-charge & \!\!$\zeta$ & \multirow{7}{*}{Continuous (inconsistent)}\\
		Battery temperature & \!\!$T_{\mathrm{b}}$ &\multirow{7}{*}{}\\
		Fuel consumption &\!\!$m_\mathrm{f}$ & \multirow{7}{*}{} \\
		Pre-DOC temperature &\!\!$T_{\mathrm{PreDOC}}$ & \multirow{7}{*}{} \\
		DOC temperature &\!\!$T_{\mathrm{DOC}}$& \multirow{7}{*}{} \\
		DPF temperature &\!\!$T_{\mathrm{DPF}}$& \multirow{7}{*}{} \\
		SCR temperature &\!\!$T_{\mathrm{SCR}}$& \multirow{7}{*}{} \\
		\hline 
		\hline 
		\textbf{Controls (4)} & \!\!\textbf{Symbol} & \textbf{Variable Type (PS3)}\\
		\hline
		Vehicle acceleration & \!\!$a$ & Continuous (consistent) \\
		Engine switch & \!\!$e_{\mathrm{cmd}}$ & Discrete (binary-valued)\\
		Gear shift command & \!\!$g_{\mathrm{cmd}}$ & Discrete (integer-valued)\\
		Torque split  & \!\!$\mu$ & Continuous (inconsistent)\\
		\hline\hline
	\end{tabular}
\end{table}

\section{Powertrain Models}
\label{sec:modeling}
All continuous states whether \textit{consistent} or \textit{inconsistent}, with their respective models, state dynamic equations, algebraic equations and constraints pertaining to them, along with relevant maps and look-up tables are listed and explained in Sections \ref{subsubsec:Bat. model}-\ref{subsubsec:AT system}. Section \ref{subsec:discdyn} is for the model dynamics of the two integer-valued control variables, namely gear shift command and engine on/off switch.
\subsection{Battery Model}
\label{subsubsec:Bat. model}
\begin{figure}[!t]
	\centering
	\subfigure[]{\includegraphics[width=0.5\linewidth]{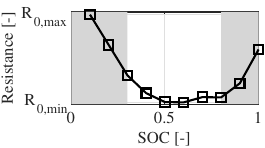}\label{fig:batt_res}}\hfil
	\subfigure[]{\includegraphics[width=0.5\linewidth]{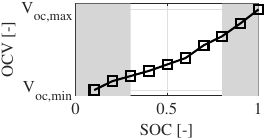}\label{fig:batt_ocv}}\hfil
	\subfigure[]{\includegraphics[width=\linewidth]{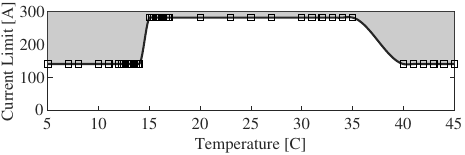}\label{fig:batt_curr}}
	\caption{(a) Internal resistance (b) Open circuit voltage (c) Temperature-dependent limit on absolute value of battery current. Infeasible regions shown in gray.}
	\label{fig:battery}
\end{figure}

An 11 kWh NMC/Graphite based battery pack of 350 V nominal voltage with 90 cells in series and 6 branches in parallel is used. There are two inconsistent state variables, battery state-of-charge (SOC) $\zeta$ and battery temperature $T_{\b}$. SOC is a dimensionless quantity between 0 and 1. A zero-th order equivalent circuit model is assumed for SOC dynamics, whereas a first order temperature model with heat addition due to Ohmic losses is used for battery thermal dynamics:
\begin{subequations}
\label{eq:batt_diffEquation}
\begin{align}
\label{SOC_diffEquation}
\dot{\zeta} &= -\dfrac{I_{\b}}{Q_{\mathrm{nom}}}; I_\b = -{\eta_{\b}}\left[\frac{V_\mathrm{oc}}{2R_0} - \sqrt{\left(\frac{V_\mathrm{oc}}{2R_0}\right)^2 - \frac{P_\b}{R_0}}\right]\!\! ,\\
\dot{T}_\b &= - \dfrac{1}{m_\b c_\b} \Big(h_{\b}A_{\b}\left(T_\b - T_{\mathrm{amb}}\right) + I_\b^2R_0\Big),
\label{BatTemp_diffEquation}
\end{align}
\end{subequations}
where, $I_\b$ is battery current, $P_\b$ is battery power, and the various constants are: $Q_\mathrm{nom}$, the battery capacity (31 Ah), $\eta_{\b}$, Coulombic efficiency (90\% for charging, 100\% for discharging), $h_{\b}$, heat transfer co-efficient due to convection, $A_{\b}$, outer battery pack surface area, $m_\b$, battery pack mass, $c_\b$, battery pack specific heat capacity, and $T_{\mathrm{amb}}$ ambient temperature (25$^\circ$ Celsius). The equivalent circuit model internal resistance, $R_0:=R_0(\zeta)$, and open circuit voltage $V_{\mathrm{oc}}:=V_{\mathrm{oc}}(\zeta)$ are based on 1-D look-up tables of SOC shown in Fig.~\ref{fig:batt_res} and \ref{fig:batt_ocv}. 
The absolute value of battery current is constrained by $I_{\mathrm{b,max}}$ which depends on temperature $T_{\b}$ as shown in Fig.~\ref{fig:batt_curr},
\begin{equation}
\label{battCurr_pathConstraint}
-I_{\mathrm{b,max}}(T_{\mathrm{b}}) \leq I_{\mathrm{b}} \leq I_{\mathrm{b,max}}(T_{\mathrm{b}}).
\end{equation}

We use linear interpolation for internal resistance and open circuit voltage signals, and spline interpolation for temperature-dependent battery current limit to retain smoothness for ease in convergence.
These dynamics are driven by the battery power, $P_\b$ which follows from energy balance:
\begin{equation}\label{eq:Pb}
P_\b = P_\m + P_\mathrm{aux},
\end{equation}
where, $P_\m$ is the mechanical power delivered to/from the electric machine, given in (\ref{eq:Pm}), and $P_\mathrm{aux}$ is a constant 5 kW accessory load on the battery. Relevant box constraint is, 
\begin{equation}
\label{batt_boxConst}
0.3 \leq \zeta \leq 0.8.
\end{equation}

Battery temperature has initial condition at ambient while SOC has charge sustaining constraint of 55\% requiring its initial and final values to be same over the time horizon,
\begin{subequations}
\label{eq:batt_boundConst}
\begin{align}
\label{SOC_boundConst}
\zeta_0 &= \zeta_T = 0.55,\\
\label{battTemp_boundConst}
T_{\mathrm{b},0} &= T_\mathrm{amb}.
\end{align}
\end{subequations}
\subsection{Vehicle Dynamics and Drivetrain}
\label{subsubsec:vehdyn}
In vehicle dynamics (eco-driving) block, there are two consistent state variables, vehicle speed $v$ and covered distance $d$, and a consistent control variable, vehicle acceleration $a$,
\begin{subequations}
\label{eq:veh_diffEquation}
\begin{align}
\label{speed_diffEquation}
\dot{v} = a,\\
\label{dist_diffEquation}
\dot{d} = v,
\end{align}
\end{subequations}

Time-varying inputs to the vehicle dynamics block are a reference drive cycle, $v_{\mathrm{org}}(t)~\forall t\in[0,T]$ that the eco-driving vehicle needs to approximately follow, and an elevation profile, $\theta_\mathrm{org}(t)~\forall t\in[0,T]$. Both these signals are shown in Fig.~\ref{fig:vehdyn}. There are two path constraints associated with this block given in (\ref{eq:vehconst}). Firstly, we restrict the eco-driven vehicle to remain within $\pm$ 5 km/h of the reference speed at all times. Secondly, a stop-at-stop constraint is imposed which forces the eco-driven vehicle to be stopped whenever the reference vehicle stops. This path constraint captures occurrences of road stop signs and red traffic lights.
\begin{equation}
\label{eq:vehconst}
|v-v_{\org}| \leq \begin{cases}
    0 & \text{if~}v_{\mathrm{org}} = 0,\\
    5 & \text{otherwise}.
\end{cases}\quad \text{[km/h]}
\end{equation}

Note that we have omitted the use of ``$(t)$'' for brevity of notation, even though the above holds at all times, and $|\cdot|$ indicates absolute value.
Initial conditions of the state variables, given below, are straightforward. Moreover, a final boundary value constraint is imposed guaranteeing that the total distance covered by eco-driven vehicle must be the same as that covered by reference.
\begin{subequations}
\label{eq:vehboxconst}
\begin{align}
\label{speed_boundConst}
v_{0}&=0,\\
\label{dist_boundConst}
d_0 &= 0,\quad d_{T}=d_{\mathrm{org},T}:=\int_{0}^{T}v_\mathrm{org}\, \d t.
\end{align}
\end{subequations}

Connected to the vehicle dynamics block is the differential and transmission block, for which the input is a gear profile $g$. A 6-speed auto transmission model is used having a constant gearbox efficiency $\eta_\mathrm{g}$ of 95\%. Longitudinal vehicle dynamics and point-mass wheel model is used for simplicity. We assume road loads of aerodynamic drag, rolling resistance, inertial drag and gradient forces acting against the supplied propulsion system power. Hence, we have the following kinematics:
\begin{subequations}\label{eq:veh_algeb}
\begin{align}
\label{eq:Fv}\begin{split}
F_{\mathrm{v}} = M_\mathrm{v}'a +\frac{c_\mathrm{d}\rho_\mathrm{a} A_\mathrm{f}}{2}v^2 + M_\mathrm{v}g_\mathrm{a}c_\mathrm{r}\cos(\theta_\mathrm{\org})\\
+ M_\mathrm{v}g_\mathrm{a}\sin (\theta_\mathrm{\org}),
\end{split}\\
\label{eq:veh_omega}
\omega &= \dfrac{\gamma_g v}{r_\mathrm{v}},\\
\label{eq:veh_alpha}
\alpha &= \dot{\omega} = \dfrac{\gamma_g a}{r_\mathrm{v}},\\
\label{eq:veh_taug}
\tau_{\mathrm{g}} &= \dfrac{F_{\mathrm{v}}r_\mathrm{v}}{\gamma_g \eta_{\mathrm{g}}^{\mathrm{sign}(F_{\mathrm{v}})}},
\end{align}
\end{subequations}
\begin{equation}
\label{eq:t_total}
\tau_{\mathrm{total}} =\! 
\begin{cases}
0 &\!\!\!\!\text{if}~ v = 0,\\
\tau_{\mathrm{g}} + \tau_{\e,\mathrm{drag}} + \alpha \left(I_\mathrm{e} + I_\mathrm{m}\right) &\!\!\!\!\text{if}~  v\neq 0\land e = 1,\\
\tau_{\mathrm{g}} + \alpha I_\mathrm{m} &\!\!\!\!\text{if}~ v\neq 0\land e = 0,
\end{cases}
\end{equation}
where, $F_\mathrm{v}$ is total traction force at wheels, $\gamma_g$ is the gear ratio for gear number $g$, $\tau_\mathrm{g}$ is the driveshaft torque after the transmission, $\tau_\mathrm{e,drag}$ is the motoring torque of the engine, i.e., rubbing friction, and $\tau_\mathrm{total}$ is the total torque that the combination of motor and engine must provide. In defining $\tau_\mathrm{total}$, we have assumed different cases depending on vehicle being stopped or not, and on engine being turned on ($e=1$) or off ($e=0$), which shows the discontinuous nature of these kinematics. Description of other constants is given in Table \ref{tab:veh_param}.
\begin{table}[!t]
	\caption{Vehicle parameters with their symbols\label{tab:veh_param}}
	\centering
	\begin{tabular}{|ll|}
		\hline
        \textbf{Symbol} & \textbf{Description}\\
		\hline
		$M_\mathrm{v}$	&	Vehicle mass\\
		$M_\mathrm{v}'$	&	Vehicle mass, scaled +10\% (for inertia)\\
		$c_\mathrm{d}$	&	Aerodynamic drag coefficient\\
		$\rho_\mathrm{a}$	&	Air density\\
		$A_\mathrm{f}$	&	Frontal area\\
		$g_\mathrm{a}$	&	Gravitational acceleration\\
		$c_\mathrm{r}$	&	Wheel rolling resistance\\
		$r_\mathrm{v}$	&	Wheel radius\\
		$\eta_\mathrm{g}$	&	Gearbox efficiency\\
		$I_\mathrm{e}$	&	Engine inertia\\
		$I_\mathrm{m}$	&	Electric machine inertia\\
		\hline
	\end{tabular}
\end{table}
Box constraints on the state and control variables are:
\begin{subequations}
\label{eq:veh_boxConst}    
\begin{align}
\label{acc_boxConst}
-2 \leq &~a \leq 1.5,&[\text{m/s}^{2}]\\
\label{speed_boxConst}
0 \leq &~v \leq 25.&[\text{m/s}]
\end{align}
\end{subequations}
\subsection{Electric Machine and Engine Models}
\label{subsubsec:Eng_Mot model}
\begin{figure}[!t]
 	\centering
 	\includegraphics*[width=0.4925\linewidth]{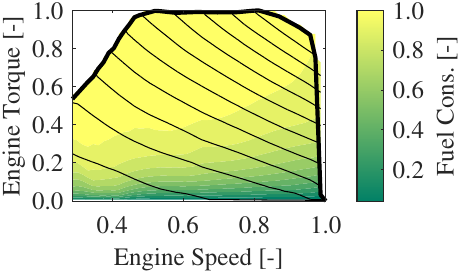}
 	\includegraphics*[width=0.4925\linewidth]{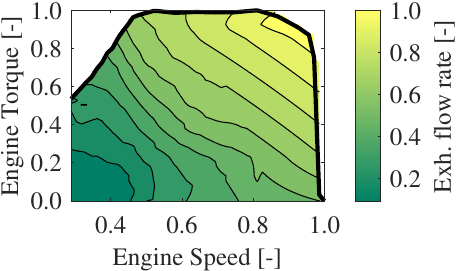}
 	\includegraphics*[width=0.4925\linewidth]{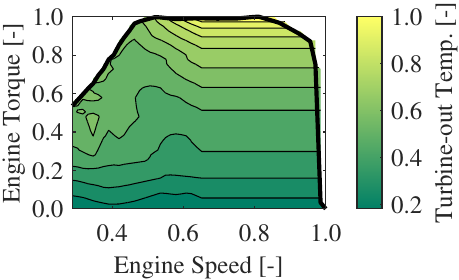}
 	\includegraphics*[width=0.4925\linewidth]{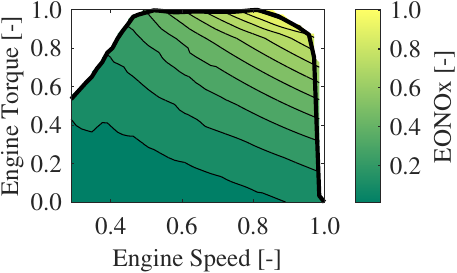}
 	\includegraphics*[width=\linewidth]{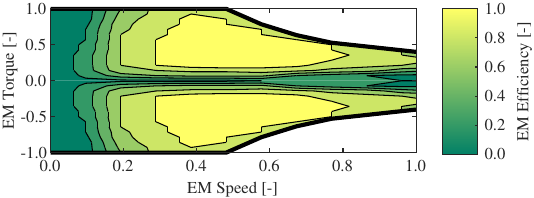}
 	\caption{Top-left: BSFC map in filled contours with fuel consumption; top-right: exhaust flow rate; center-left: turbine-out temperature; center-right: engine-out NOx; bottom: electric machine efficiency. Maps are normalized for confidentiality.\label{fig:eng_mot maps}}
\end{figure}

A 90 kW electric machine (EM) is used operating in continuous mode, modeled as a lumped mechanical-electrical conversion efficiency map, $\eta_\m(\omega,\tau_\m)$ of the operating shaft speed $\omega$ and EM torque $\tau_{\m}$. For the internal combustion engine (ICE), mean-value model of a 270 horsepower diesel engine is considered, which is down-scaled to 220 horsepower rating. ICE maps for fuel consumption $\dot{m}_\f$, exhaust flow rate $\dot{m}_{\text{exh}}$, turbine-out temperature $T_{\text{TOT}}$, and engine-out NOx $\dot{m}_{\e}$ are 2-D look-up tables of shaft speed $\omega$ and engine torque $\tau_{\e}$:
\begin{subequations}\begin{align}\label{eq:mfdot}
\dot{m}_\f &= \mathcal{F}(\omega,\tau_\e),\\\label{eq:mfexh}
\dot{m}_{\text{exh}} &= \mathcal{M}(\omega,\tau_\e),\\\label{eq:tot}
T_{\text{TOT}} &= \mathcal{T}(\omega,\tau_\e),\\\label{eq:mfe}
\dot{m}_{\e} &= \mathcal{N}(\omega,\tau_\e).
\end{align}\end{subequations}
These normalized maps are depicted in Fig.~\ref{fig:eng_mot maps}.
When the driveshaft speed is below engine idle these four signals take some minimum values. If engine is turned off, the three flow rate signals take value of zero, and turbine-out temperature takes idling temperature value. The torques of engine, $\tau_{\e}$ and EM, $\tau_{\m}$ relate algebraically with the demand torque, $\tau_{\mathrm{total}}$ through the inconsistent torque split control, $\mu$, differently in traction and braking phases of the drive cycle:
\begin{equation}\label{eq:traction}
\underset{\tau_{\mathrm{total}}\geq 0}{\text{Traction}}\begin{cases}
\tau_{\e} =\! \begin{cases}
\left(1-\mu\right) \!\tau_{\mathrm{total}} &\!\!\!\!\!\text{if}~\omega \geq \omega_{\mathrm{idle}}\land e=1,\\
0 &\!\!\!\!\! \text{if}~\omega < \omega_{\mathrm{idle}}\lor e=0,
\end{cases}\\
\tau_{\m} =\! \begin{cases}
\mu \tau_{\mathrm{total}} &\text{if}~\omega \geq \omega_{\mathrm{idle}}\land e=1,\\
\tau_{\mathrm{total}}&\text{if}~\omega < \omega_{\mathrm{idle}}\lor e=0,
\end{cases}
\end{cases}
\end{equation}
\begin{equation}\label{eq:braking}
\underset{\tau_{\mathrm{total}}< 0}{\text{Braking}}\begin{cases}
\tau_{\e} = 0,\\
\tau_{\m} = \max\{\tau_{\mathrm{total}},\tau_{\m,\min}\}.
\end{cases}
\end{equation}

When the vehicle is braking, i.e. $\tau_{\mathrm{total}} < 0$, we assume that EM operates at maximum recuperation energy to charge the battery which is illustrated as $\max\{\tau_{\mathrm{total}},\tau_{\m,\min}\}$. It can be noticed that the torque split control variable becomes free while braking or whenever the engine is off. Note that engine drag is accounted for by adding it in the demand torque expression $\tau_{\mathrm{total}}$ (\ref{eq:t_total}). Through experimentation with the solver, we learnt that defining $m_\mathrm{f}$ as a state variable, as opposed to a dependent signal, aids in convergence. Torque split can take negative values (engine charges battery) as well as positive values (engine and battery joint propulsion):
\begin{subequations}\label{eq:propboxbound}
\begin{align}\label{eq:mu_box}
-1 &\leq \mu \leq 1,\\\label{eq:mf_box}
0 &\leq m_\f,
\end{align}
\end{subequations}
\begin{equation}
\label{eq:mfbound}
m_{\f,0} = 0.
\end{equation}

The mechanical power delivered to/from electric machine, $P_\m$, that discharges/charges the battery, algebraically relates with electric machine (EM) torque through EM efficiency,
\begin{equation}\label{eq:Pm}
P_\m = \frac{\omega \tau_{\m}}{\eta_\m^{\mathrm{sign}(\tau_\m)}}.
\end{equation}
 
The relevant path constraints are given below: the engine and EM torques are limited at their minimum and maximum curves, which are given by 1-D look-up tables (shown as black curves in Fig.~\ref{fig:eng_mot maps}) of shaft speed which is constrained by maximum allowable engine/EM speed:
\begin{subequations}
\label{eq:pathEMICE}
\begin{gather}
\label{MotTorq_PathConst}
\tau_{\m,\min} \leq \tau_{\m} \leq \tau_{\m,\max},\\
\label{EngTorq3_PathConst}
\tau_{\e,\min} \leq \tau_{\e} \leq \tau_{\e,\max}\iff e = 1,\\
\label{shaftSpeed_PathConst}
0 \leq \omega \leq \omega_{\max}:=\min \{\omega_{\e,\max},\omega_{\m,\max}\}.
\end{gather}
\end{subequations}
\subsection{After-treatment Model}
\label{subsubsec:AT system} 
\begin{figure}[!t]
 	\centering
 	\includegraphics*[width=\linewidth]
  {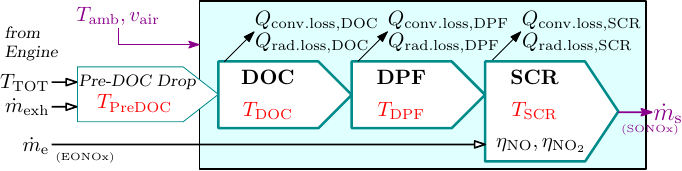}
 	\caption{Block diagram of thermal and emissions model in after-treatment system. Temperature state variables are in red.\label{fig:aft_diag}}
\end{figure}

The after-treatment system consists of four-stage thermal dynamics. Firstly, it has a calibrated first-order model of temperature drop for the exhaust gases due to the pipe connecting engine exhaust and after-treatment system. Then, consecutive first-order validated models of the catalysts, Diesel Oxidation Catalyst (DOC), then Diesel Particulate Filter (DPF), and finally, Selective Catalytic Reduction (SCR) are assumed. In total there are four inconsistent states associated with the after-treatment system, which are exhaust gas temperatures in Pre-DOC pipe, DOC, DPF and SCR. Pressure variation is not modeled. Ambient losses are assumed to be only due to convection and radiation. Pictorially, the thermal flow is shown in Fig. \ref{fig:aft_diag}.
State dynamic equations are given by:
\begin{subequations}
\label{eq:aftdiff}
\begin{align}
\label{Pre-DOC_diffEquation}
\dot{T}_{\text{PreDOC}} &= 0.042 \dot{m}_{\text{exh}}(T_{\text{TOT}} - T_{\text{PreDOC}}),\\
Q_{\text{in,}(\cdot)} & = c_{\text{p,air}}\dot{m}_{\text{exh}}(T_{(*)} - T_{(\cdot)}),\\
Q_{\text{conv.loss,}(\cdot)} & = h_{(\cdot)}A_{(\cdot)}(T_{\text{amb}} - T_{(\cdot)}),\\
Q_{\text{rad.loss,}(\cdot)} & = \epsilon\sigma_{(\cdot)}A_{(\cdot)}(T_{\text{amb}}^{4} - T_{(\cdot)}^{4}),\\
\dot{T}_{(\cdot)} & = \frac{Q_{\text{in,}(\cdot)} + Q_{\text{conv.loss,}(\cdot)} + Q_{\text{rad.loss,}(\cdot)}}{m_{(\cdot)}c_{\text{p,}(\cdot)}},
\label{ATtemp_diffEquation}
\end{align}\end{subequations}
where, $(\cdot) \in \text{\{DOC, DPF, SCR\}}$, $T_{(*)}$ is the temperature of the previous stage, $Q_{\text{in},(\cdot)}$ is the energy entering the catalyst $(\cdot)$, $Q_{\text{conv.loss},(\cdot)}$ is the loss of energy due to convection, and $Q_{\text{rad.loss},(\cdot)}$ is the loss due to radiation. The specific heat values of the DOC, $c_{\mathrm{p},\mathrm{DOC}}$ and DPF, $c_{\mathrm{p},\mathrm{DPF}}$ are constants, while that of SCR, $c_{\mathrm{p},\mathrm{SCR}}$ is a linearly interpolated look-up table dependent on SCR temperature. The heat transfer co-efficients $h_{(\cdot)}$ are nonlinear functions of air speed (assumed equal to vehicle speed), ambient temperature, respective catalyst lengths and catalyst external heat factor constants. Other constants are catalyst surface area $A_{(\cdot)}$, catalyst mass $m_{(\cdot)}$, catalyst external emissivity $\sigma_{(\cdot)}$, specific heat capacity of air $c_{\mathrm{p},\mathrm{air}}$, and the Stefan-Boltzmann constant $\epsilon$. Initial conditions for all four temperatures are set at ambient,
\begin{equation}
    \label{eq:boundTemps}
    T_\mathrm{PreDOC,0}=T_\mathrm{DOC,0}=T_\mathrm{DPF,0}=T_\mathrm{SCR,0}=T_\mathrm{amb}.
\end{equation}

\begin{figure}[!t]
 	\centering
 	\includegraphics*[width=0.49\linewidth]{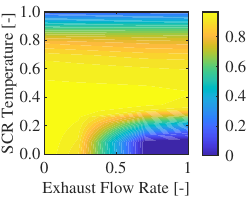}
 	\includegraphics*[width=0.49\linewidth]{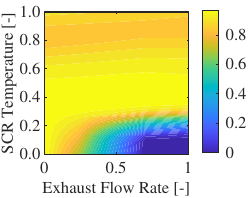}
 	\caption{Normalized 2-D Maps of SCR block's NOx conversion efficiencies: $\eta_{\mathrm{NO}}$ (left) and $\eta_{\mathrm{NO}_2}$ (right).\label{fig:conv_eff maps}}
 \end{figure}

The emissions model consists of SCR's NO and NO$_2$ conversion efficiency maps which are linearly interpolated dependent on SCR temperature and exhaust flow rate, $\eta_{\mathrm{NO}}(T_{\mathrm{SCR}},\dot{m}_{\mathrm{exh}})$ and $\eta_{\mathrm{NO}_2}(T_{\mathrm{SCR}},\dot{m}_{\mathrm{exh}})$ | see Fig.~\ref{fig:conv_eff maps}. Exhaust gases density is assumed equal to air density. System-out NOx, $\dot{m}_\mathrm{s}$ is the product of engine-out NOx and the conversion (in)efficiencies assuming equal NO and NO$_2$ molecules,
\begin{equation}
\label{eq:noxeff}
\dot{m}_\mathrm{s} = \frac{\dot{m}_\mathrm{e}}{2} (1-\eta_{\mathrm{NO}}) + \frac{\dot{m}_\mathrm{e}}{2} (1-\eta_{\mathrm{NO}_2}).
\end{equation}
\subsection{Gear and Engine On/Off Switch Models}
\label{subsec:discdyn}
Gear command discrete control signal, $g_{\mathrm{cmd}}\in \{-5,\dots,-1,0,1,\dots,5\}$, performs instantaneous gear shifts impacting the gear number selection, $g \in \{1,2,3,4,5,6\}$ discrete state in its difference equation,
\begin{equation}
\label{gear_equation_1}
g(k+1) - g(k) \!= \!\begin{cases}
g_{\mathrm{cmd}} &\!\!\!\text{if} ~ \sigma_{g}(k) > L,\\
0 &\!\!\!\text{otherwise},
\end{cases}
\end{equation}
where $k$ is the discretized time step,  $L=3$ seconds is minimum allowed dwell-time between consecutive shifts, and $\sigma_{g}(k)$ is dwell-time counter state variable that keeps track of the time elapsed since last gear shift, governed by,
\begin{equation}
\label{gear_equation_2}
\sigma_{g}(k+1) - \sigma_{g}(k) \!= \!\begin{cases}
-(L\! + \!1) &\!\!\!\text{if} ~ g(k+1) - g(k) = g_{\mathrm{cmd}},\\
1 &\!\!\!\text{else~if} ~ \sigma_{g}(k) \leq L,\\
0 &\!\!\!\text{otherwise}.
\end{cases}
\end{equation}
Similar to the gears are the dynamics of engine on/off discrete state variable, $e \in \{0,1\}$ using engine on/off switch control, $e_{\mathrm{cmd}}\in \{-1,0,1\}$ and dwell-time counter $\sigma_{e}$ with minimum dwell-time of $L_{\e}=2$ seconds,
\begin{equation}
\label{eng_equation_1}
e(k+1) - e(k)\! = \!\begin{cases}
e_{\mathrm{cmd}} &\!\!\!\text{if} ~ \sigma_{e}(k) > L_{\e},\\
0 &\!\!\!\text{otherwise},
\end{cases}
\end{equation}
\begin{equation}
\label{eng_equation_2}
\sigma_{e}(k+1) - \sigma_{e}(k)\! =\! \begin{cases}
-(L_{\e}\! +\! 1) &\!\!\!\text{if} ~ e(k+1) - e(k) = e_{\mathrm{cmd}},\\
1 &\!\!\!\text{else~if} ~ \sigma_{e}(k) \leq L_{\e},\\
0 &\!\!\!\text{otherwise}.
\end{cases}
\end{equation}

\section{PS3 Implementation}
\label{sec:Implementation}
The case-study optimal control problem we aim to solve seeks to minimize cost function $J$ given in (\ref{eq:cost_func}) with system governed by 13 states and 4 controls listed in Table \ref{tab:S&C} subject to all differential equations, path, boundary, and box constraints given in (\ref{eq:batt_diffEquation})-(\ref{eng_equation_2}). Given the mixed-integer nonlinear nature of the problem we use the pseudo-spectral collocation based PS3 algorithm \cite{paper1citation} for its solution. The continuous-time problem is discretized into $N:=1200$ control intervals of $1$ second step size each and is solved in three steps of consecutive numerical programs. For complete algorithmic details we refer the reader to \cite{paper1citation}.
The following sections explain the three steps of the PS3 algorithm implementation to solve our case-study problem.

\subsection{Step-1: Solving Relaxed NLP}
\label{subsec:S1}
The NLP solved in PS3's first step has nine state variables: $\zeta$, $v$, $d$, $m_{\f}$, $T_{\b}$, $T_{\text{PreDOC}}$, $T_{\text{DOC}}$, $T_{\text{DPF}}$, and $T_{\text{SCR}}$, and has four control variables: $\mu$, $\tilde{g}$, $a$, and $\tilde{e}$, where $\tilde{g}$ and $\tilde{e}$ are relaxed control equivalents of gear number and engine on/off states,
\begin{subequations}
\label{eq:boxrel}
\begin{align}
\label{eq:gearrel}
0.5 &\leq \tilde{g}\leq 6.5,\\
\label{eq:engrel}
0 &\leq \tilde{e}\leq 1.
\end{align}
\end{subequations}

Since the key aspect of step-1 is the relaxation of integer-valued variables, some of the algebraic relationships and path constraints were modified to accommodate the relaxed nature of gear number and engine switch. For example, demand torque instead of being calculated by (\ref{eq:t_total}) is done so by:
\begin{equation}\label{eq:t_total_rel}
\tau_{\mathrm{total}} =\! 
\begin{cases}
0 &\!\!\text{if}~ v = 0,\\
\tau_{\mathrm{g}} + \tilde{e}\tau_{\e,\mathrm{drag}} + \alpha \left(\tilde{e}I_\mathrm{e} + I_\mathrm{m}\right) &\!\!\text{otherwise}.
\end{cases}
\end{equation}
Likewise, dependence of $e$ in calculation of engine and EM torques during traction (\ref{eq:traction}) is ignored, and the path constraint limiting the engine torque in (\ref{EngTorq3_PathConst}) is instead formulated as follows:
\begin{align}
\label{engine_PathConst}
\tilde{e}\cdot\tau_{\e,\min} \leq \tau_{\e} \leq \tilde{e}\cdot\tau_{\e,\max}.
\end{align}
For fractional gear choices, the gear ratio $\gamma_g$ is assumed to be a linearly interpolated function of $\tilde{g}$.
To sum up, the step-1 relaxed NLP is defined by,
\begin{itemize}
\item cost function (\ref{eq:cost_func}),
\item ordinary differential equations (\ref{eq:batt_diffEquation}), (\ref{eq:veh_diffEquation}), (\ref{eq:mfdot}), (\ref{eq:aftdiff}),
\item initial- and final-value constraints (\ref{eq:batt_boundConst}), (\ref{eq:vehboxconst}), (\ref{eq:mfbound}),~(\ref{eq:boundTemps}),
\item box constraints (\ref{batt_boxConst}),~(\ref{eq:veh_boxConst}),~(\ref{eq:propboxbound}),~(\ref{eq:boxrel}),
\item and path constraints (\ref{battCurr_pathConstraint}),~(\ref{eq:vehconst}),~(\ref{MotTorq_PathConst}),~(\ref{shaftSpeed_PathConst}),~(\ref{engine_PathConst}).
\end{itemize}

\subsection{Step-2: Solving Integer States and Controls}
\label{subsec:S2}
Once step-1 is solved, we obtain the optimal trajectories of consistent variables $(v, d, a)$ and the trajectories of the relaxed discrete variables $(\tilde{g},\tilde{e})$. From this point onward, the trajectories of consistent variables $(v, d, a)$ are held known and fixed. Step-2 of the PS3 algorithm is about finding the optimal integer trajectories $(g,e)$ from the relaxed solutions that satisfy dwell-time constraints by solving a mixed-integer quadratic program (MIQP). Effectively, it solves an optimal control problem of four discrete states $(g,e,\sigma_g,\sigma_e)$ with difference equations (\ref{gear_equation_1})-(\ref{eng_equation_2}) and two discrete controls $(g_\mathrm{cmd},e_\mathrm{cmd})$ subject to relevant constraints.
However, by reformulation as an MIQP we have $N=1200$ engine on/off related and $6\cdot N=7200$ gear number related binary variables. The binary gear trajectory is denoted by $b_j(k)\in \{0$ ($j$-th gear inactive)$,1$ ($j$-th gear active)$\}$. Similarly, the relaxed equivalent of binary gear trajectory is denoted using $r'_j(k)\in [0,1]$ which is known directly from $\tilde{g}(k)$ trajectory.

Before the MIQP is defined, we start off from the optimal trajectories of consistent variables, $v$ and $a$ to arrive at all possible shaft speed and shaft angular acceleration values for the 6 gears at every time step. Naturally, not all gears will always be feasible in the complete drive cycle due to violation of the maximum shaft speed constraint (\ref{shaftSpeed_PathConst}). Another reason for infeasibility of a gear at a given time is when the corresponding maximum torque constraint (\ref{MotTorq_PathConst})(\ref{EngTorq3_PathConst}) is violated, which also depends on the engine on/off status. To capture these path constraints (\ref{eq:pathEMICE}), we arrive at two gear-feasibility binary matrices $B_{0}$ (when $e=0$) and $B_{1}$ (when $e=1$), each of size $N\times 6$. Other path constraints (\ref{battCurr_pathConstraint}) and (\ref{eq:vehconst}) of the original control problem are irrelevant. For the $k$-th time step (out of $N$ steps), and $j$-th gear number,
\begin{align*}
    B_{0,j}(k) &:= \begin{cases}
1\quad\text{if~}\tau_{\mathrm{total},j}(k) \leq \tau_{\m,\max,j}(k) \land \omega_j(k) \leq \omega_{\max},\\
0\quad \text{otherwise},
\end{cases}\\
B_{1,j}(k) &:= \begin{cases}
\begin{aligned}1\quad \text{if~}\tau_{\mathrm{total},j}(k) \leq \left(\tau_{\e,\max,j}(k)+\tau_{\m,\max,j}(k)\right) \\\land ~\omega_j(k) \leq \omega_{\max},\end{aligned}\\
0\quad \text{otherwise}.
\end{cases}
\end{align*}

Once the two gear-feasibility matrices are determined, the following MIQP is formulated and solved:
\begin{align*}
\min_{e(k),b_j(k)}~
&\sum_{k=1}^{N}\!\left(\left(e(k)-\tilde{e}(k)\right)^2+\sum_{j=1}^{6}\left(b_{j}(k)-r_{j}'(k)\right)^2\right),
\\
\mathrm{s.t.}~
&\textit{One-Gear-At-A-Time Constraint~}\forall k:\\
&
\sum_{j=1}^{6} b_{j}(k) = 1,
\\
&\textit{Feasible Gear Selection Constraint~}\forall k~\forall j:\\
&
0\leq b_{j}(k) \leq \begin{cases}
B_{0,j}(k)\qquad \mathrm{if~}e(k)=0,\\
B_{1,j}(k)\qquad \mathrm{if~}e(k)=1,
\end{cases}
\\
&\textit{Minimum Dwell-Time Constraints~}\forall k~\forall j:\\
&\forall i\in \left\{k,k+1,\cdots,k+L\right\}:\\
&~\qquad b_{j}(k) - b_{j}(k-1) \leq b_{j}(i),\\
&~\qquad b_{j}(k-1) - b_{j}(k) \leq 1 - b_{j}(i),\\
&\forall i_\e\in \left\{k,k+1,\cdots,k+L_\e\right\}:\\
&~\qquad e(k) - e(k-1) \leq e(i_\e),\\
&~\qquad e(k-1) - e(k) \leq 1 - e(i_\e),
\end{align*}
where, $L=3$ seconds and $L_\e=2$ seconds are the minimum dwell-time durations for gear shifts and engine switches respectively.
Notice that the \textit{feasible gear selection constraint} is an ``indicator'' constraint because the upper bound imposed on the optimization variable $b_j(k)$ is either of the two pre-determined values $B_{0,j}(k)$ or $B_{1,j}(k)$, but the choice is governed by the value of another optimization variable $e(k)$. It is common knowledge in integer programming that indicator constraints can be written as linear inequality constraints. Hence our step-2 problem is a mixed-integer quadratic programming problem as it only has linear constraints on the optimization variables with a quadratic objective function. It is solved using MIQP solver, Gurobi \cite{gurobi}, with solution time under 10 seconds.

As a result of solving the above described MIQP, we obtain the optimal discrete trajectories of gear number ${g}(k)$ and engine state ${e}(k)$, which are then fixed and used in step-3 to solve for the remaining inconsistent variables.

\subsection{Step-3: Solving for the Inconsistent Variables}
\label{subsec:S3}
The NLP solved in PS3's third step has all seven inconsistent state variables: $\zeta$, $m_{\f}$, $T_{\b}$, $T_{\text{PreDOC}}$, $T_{\text{DOC}}$, $T_{\text{DPF}}$, and $T_{\text{SCR}}$, and single inconsistent control variables: $\mu$.
The optimized vehicle speed $v$ and acceleration $a$ profiles from step-1, and the integer gear $g$ and engine on/off $e$ profiles from step-2 become known inputs to this step. Outcome of this step are the optimal trajectories of all inconsistent variables, completing the solution. The NLP is defined by, 
\begin{itemize}
\item cost function (\ref{eq:cost_func}) which is retained from Section~\ref{subsec:S1},
\item ordinary differential equations (\ref{eq:batt_diffEquation}), (\ref{eq:mfdot}), (\ref{eq:aftdiff}),
\item initial- and final-value constraints (\ref{eq:batt_boundConst}), (\ref{eq:mfbound}),~(\ref{eq:boundTemps}),
\item box constraints (\ref{batt_boxConst}),~(\ref{eq:propboxbound}),
\item and path constraints (\ref{battCurr_pathConstraint}),~(\ref{eq:pathEMICE}).
\end{itemize}
\subsection{Solver Specifics and Initial Guess}
The polynomial degree for pseudo-spectral collocation points was set to $5$ to take full advantage of Legendre-Gauss-Radau collocation for handling stiffness.
Interior point algorithm IPOPT \cite{wachter2006implementation} with Harwell Subroutine Library MA97 linear solver \cite{hsl2007collection} for solving nonlinear programs
is used. With regards to the associated tuning effort and initial guess used, (a) normalization of fuel and emissions terms in the cost function followed by numerical scaling was adjusted in steps 1 and 3 to make sure that the final numerical value remains within $0$ and $1$; (b) IPOPT error tolerances were usually set as \texttt{tol}$=10^{-5}$ and \texttt{acceptable\_tol}$=10^{-1}$; (c)
the initial guess of the four control signals in step 1 was generated using a combination of the baseline solution and rule-based strategy, which was run by \texttt{YopSimulator} class \cite{leek2016optimal} to obtain initial guesses of all state variable trajectories. In particular, the initial guess of the torque split control was set to be na\"ive, a straight line starting the drive cycle at $\bar{\mu}_{0}=1$ (battery-only mode) and ending at $\bar{\mu}_{T}=0$ (engine-only mode). Furthermore, IPOPT's options of \texttt{fixed\_variable\_treatment=`relax\_bounds'} and \texttt{bound\_relax\_factor}$=10^{-4}$ were used for ease in convergence. Warm start option is enabled in step 3 which takes the relaxed solution of step 1 as its initial guess.
\section{Results and Analysis}
\label{sec:Results}
The following sections outline the results of experiments we performed using our approach. We first present the Pareto-front study and compare overall fuel consumption and NOx emissions numbers in the three problems. Then we analyze trajectories of various dynamic state, control and other signals comparing three problem cases (\textit{Fuel problem}, \textit{Emissions problem} and \textit{Fuel \& Emissions problem}) to establish comprehensiveness and energy footprint impact, which culminates with benchmarking energy analysis of all powertrain components. Computation time for these problems are of the order of 45-75 minutes for each problem run.
\subsection{Pareto-optimal Study}
\label{subsec:pareto study}

\begin{figure}[!t]
	\centering
	\includegraphics*[width=\linewidth]{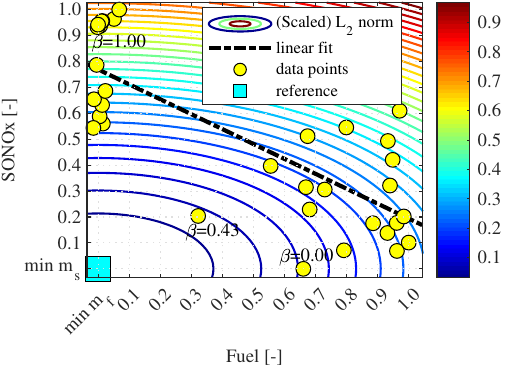}
	\caption{Pareto-front study showing data points for various values of $\beta$, a linear regression fit $y(x)=-3.88x+16.2$, and Euclidean distance contours from a reference point (axes are normalized between $1.0$ and minimum $m_\f$ or $m_\mathrm{s}$).\label{fig:pareto}}
\end{figure}

Various values of the weighting factor $\beta$ in (\ref{eq:cost_func}) were chosen to solve Fuel \& Emissions problem, while keeping solver options, initial guess, error tolerances and objective scaling the same. Resulting values of total fuel consumed and NOx emissions are shown in Fig.~\ref{fig:pareto}. Wide spread of values owes to the fact that the algorithm may converge to local minima. The two objective function terms have different sensitivities to $\beta$ and thus the two axes are scaled for finding Pareto-optimal point by Euclidean norm. As a result, $\beta=0.43$ is chosen as the best compromise between fuel and emissions minimization. For faster computation time in the Pareto-front study, we use Radau collocation of degree one. The two extreme data points, $\beta=1.00$ and $\beta=0.00$ are annotated in the figure, and so is the Pareto-optimal $\beta=0.43$.
\subsection{Fuel Consumption and System-Out NOx Emission}
\label{subsec:Results_overall}
For the three cases of $\beta=1$, $\beta=0.43$, and $\beta=0$ we present the primary overall metrics of respective obtained optimal solutions in Table \ref{tab:overall result}. 
To compare our results with a baseline, we solved a simpler optimization problem having only a single control variable, the torque split, on the same but coarsely modeled powertrain. Baseline operates on the reference drive cycle, assumes engine to be turned off whenever vehicle is stopped and gear sequence to be predefined by speed-dependent rules. A case minimizing only cumulative fuel using Dynamic Programming, and another minimizing only cumulative NOx emissions using PS3 were considered for the baseline. In Table \ref{tab:overall result} we observe that the Pareto-optimal result, i.e., Fuel \& Emissions problem, has 7\% reduction in fuel consumption, and 29\% reduction in NOx emissions compared to the best baseline solutions. Net energy demand at the wheels, which is an outcome of eco-driving control of vehicle speed can be observed to have reduced by 6\% compared to the baseline. Note that due to a hard constraint set up, the total distance covered and total trip time exactly matches with the reference drive cycle for all three problems.

\begin{figure}[!t]
\centering
\includegraphics*[width=\linewidth]{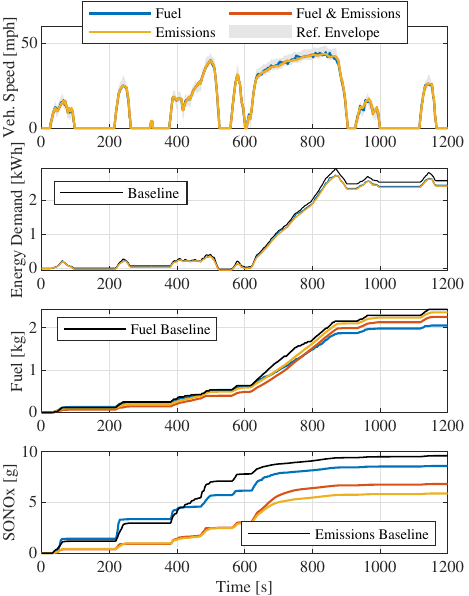}
\caption{Vehicle speeds, fuel consumption and system-out NOx emission for the three problems and baselines.\label{fig:signals1}}
\end{figure}

Fig.~\ref{fig:signals1} shows the trajectories of vehicle speed profile (which is one of the many optimized variables and varies within $\pm 5$ km/h envelope of reference), powertrain energy demand (a direct consequence of eco-driven vehicle speed), and cumulative signals of fuel consumed and system-out NOx emitted. 
\begin{table}[!t]
	\caption{Problem-wise overall fuel consumption, system-out NOx emission and net powertrain energy demand at wheels\label{tab:overall result}}
	\centering
	\begin{tabular}
 {m{0.23\linewidth}m{0.18\linewidth}m{0.21\linewidth}m{0.17\linewidth}}
		\hline
		\textbf{Problem Name} & \textbf{Fuel}, $m_\f$ [kg] & \textbf{SONOx}, $m_\mathrm{s}$ [g] & \textbf{Energy Demand} [kWh]\\
		\hline\hline
            \multicolumn{4}{c}{\textit{Baseline}}\\
		\hline
         Min. Fuel   & {2.43} & | & {2.569}\\
         Min. Emissions   & | & {9.59} & {2.569}\\
		\hline
            \multicolumn{4}{c}{\textit{PS3 optimized}}\\
		\hline
		Fuel & 2.05 \footnotesize{($\blacktriangledown 15\%$)} & 8.58 \footnotesize{($\blacktriangledown 10\%$)} & 2.440 \footnotesize{($\blacktriangledown 5\%$)}\\
		Fuel \& Emissions & 2.26 \footnotesize{($\blacktriangledown 7\%$)} & 6.81 \footnotesize{($\blacktriangledown 29\%$)} & 2.411 \footnotesize{($\blacktriangledown 6\%$)}  \\
		Emissions & 2.36 \footnotesize{($\blacktriangledown 3\%$)} & 5.87 \footnotesize{($\blacktriangledown 38\%$)}& 2.414 \footnotesize{($\blacktriangledown 6\%$)}   \\
		\hline
	\end{tabular}
\end{table}


\subsection{Analysis of State and Control Signal Trajectories}
\label{subsec:resultsignals}
\begin{figure*}[th!]
\centering
 \includegraphics*[width=\linewidth]{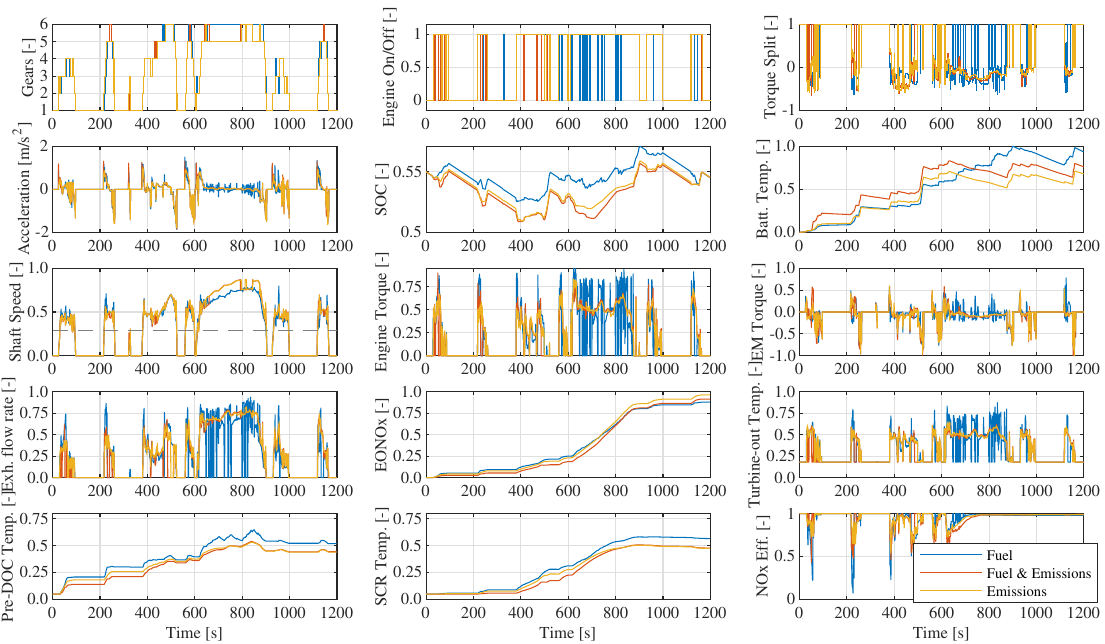}
\caption{Resultant optimal trajectories of various signals (axes normalized for data confidentiality)\label{fig:signals2}}
\end{figure*}
Fig.~\ref{fig:signals2} shows resultant trajectories of various time-series signals obtained after solving the three optimal control problems. To take into account electric energy usage, a charge-sustaining constraint on battery SOC to have its initial and final values at 55\% is imposed. Also, minimum dwell-time constraints on gear shifting and engine on/off switching are imposed as well to avoid chattering and improve drivability.

For fuel efficiency, we can observe that the solution of Fuel problem prefers frequent engine on/offs and higher gear in the high power maneuver at 600-900 seconds of the drive cycle, which is where most fuel is saved. The Emissions and Fuel \& Emissions problems tend to take lower gear and keep engine on in this maneuver, and raise the catalyst temperatures for better pollutant reduction. However, initial 300 seconds are where the Fuel problem consumes most fuel by operating at higher engine torque to charge up the battery, where the Fuel \& Emissions problem saves more by turning engine off frequently. Cumulative fuel plot in Fig.~\ref{fig:signals1} verifies the trend.

As for behavior with respect to NOx emissions, various temperatures and other engine-out signals are depicted in the same figure. Even though temperature of aftertreatment system's SCR block has the primary role in efficient conversion of NOx pollutants, yet highest SCR temperature does not necessarily guarantee overall reduction in system-out NOx because of dependence on other terms of engine exhaust flow rate and engine-out NOx. In the plots we can observe that the SCR temperature is usually always lower for Fuel \& Emissions problem compared to the other two solutions, but because of engine-out NOx and exhaust mass flow rate also being consequently lower (especially in earlier cold part of drive cycle), the SCR's NOx conversion efficiencies are higher on average. Specifically, we see dips in NOx efficiency in earlier half where exhaust flow rates shoot up due to high power demand from engine. Thus, we see Fuel \& Emissions problem doing relatively better emission reduction compared to Fuel problem, despite having slightly lower SCR temperature. Cumulative SONOx plot in Fig.~\ref{fig:signals1} verifies the trend.

\begin{figure}[!t]
	\centering
	\subfigure[]{\includegraphics*[width=\linewidth]{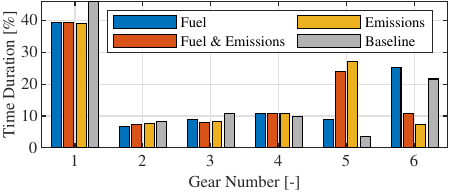}\label{fig:gearbars}}\hfil
	\subfigure[]{\includegraphics*[width=0.5\linewidth]{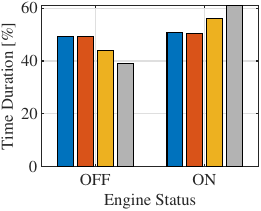}\label{fig:engbars}}\hfil
	\subfigure[]{\includegraphics*[width=0.5\linewidth]{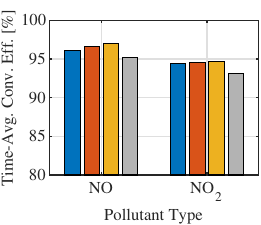}\label{fig:noxbars}}
	\caption{(a) Gear durations (b) Engine on/off durations (c) Time-averaged SCR conversion efficiencies.\label{fig:barcharts}}
\end{figure}

Fig.~\ref{fig:barcharts} shows bar charts of cumulative behaviors of the gear selection control, engine on/off control and the performance of average NOx conversion efficiencies. We observe that the Fuel \& Emissions problem solution tends to prefer lower gears overall compared to the Fuel problem in order to keep lower NOx emissions at the cost of higher fuel. On the other hand, the Emissions problem has longer engine on duration (56\%) compared to the other two (50.6\% and 50.7\%). Thus, keeping engine off for long, especially in the first half of the duty cycle, allows the Fuel \& Emissions problem to save more on fuel compared to Emissions problem. Time-averaged NOx conversion efficiencies conform to the objective functions of the three respective problems. Note that in our aftertreatment emissions model we have assumed equal ratio of NO and NO$_2$ molecules, and have not considered ammonia storage nor catalytic pressures. Baseline time-duration and efficiency percentages are given for reference comparison.

\subsection{Benchmarking Energy Analysis}
\label{subsec: Benchmarking}
\begin{figure*}[!t]
	\centering
	 \includegraphics[width=\linewidth]
  {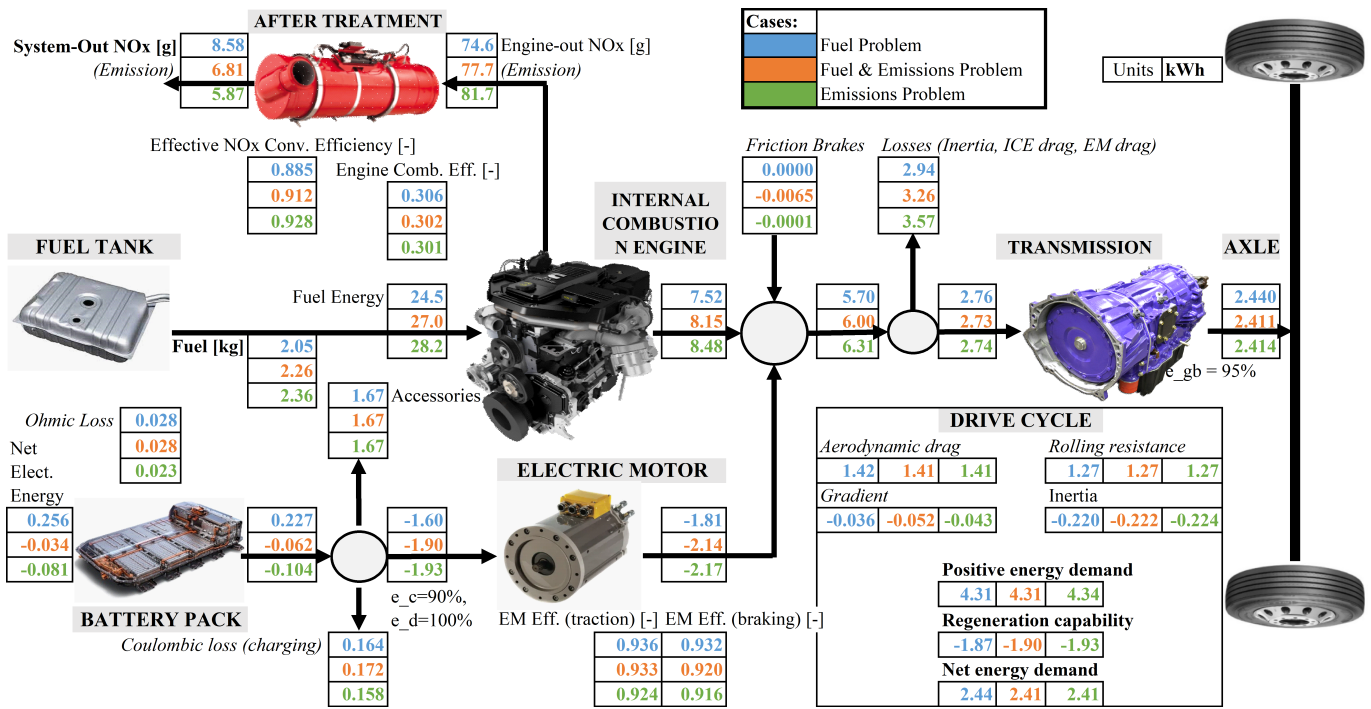}
	\caption{Comparative analysis of net energy flow in the three cases.\label{fig:energy}}
\end{figure*}
As previously established in our prequel, this problem cannot be easily solved using the well-known global optimization benchmark algorithm, Dynamic Programming, because of its curse of dimensionality for problems of such a large number of variables. For our results to serve as benchmark, we present a cumulative energy analysis. In Fig. \ref{fig:energy} overall net-energy flow numbers are shown between various powertrain components for the three problem cases. All boxes showing net energy are in kWh. Other terms, such as efficiency, fuel, and emissions are in their respective units as shown.

We can observe certain trends in Fig. \ref{fig:energy}.
In the Fuel problem, we see that engine's combustion efficiency (30.6\%) is the highest, and net electrical energy drawn from the battery is positive, i.e., the battery is providing energy overall. Both of these observations conform to the objective of Fuel problem and strengthen the optimality of its results. The Emissions problem, is opposite in these aspects. However, Emissions problem has the largest positive energy demand (4.34 kWh) as well as regeneration capability (-1.93 kWh) in its eco-driven drive cycle. High positive energy demand leads to high engine provided energy, which indirectly improves the effective NOx conversion efficiency to 92.8\% compared to the Fuel problem's 88.5\%. High regeneration capability helps in more battery charging operation (-0.081 kWh), lesser Ohmic loss (0.023 kWh) and the least Coulombic loss while charging (0.158 kWh). Finally, the Fuel \& Emissions problem mostly has values (orange) within the other two extremes (blue and green). This verifies the appropriate balance in Pareto-optimality of fuel consumption and NOx pollutant emissions. Not only are overall fuel and emission numbers being traded off, but the component-wise efficiency, energy delivered and consumed, as well as mechanical losses are also the result of this trade-off. When comparing performance of eco-driving control, the Fuel \& Emissions problem has the lowest net energy demanded at the wheels (2.411 kWh), supporting Pareto-optimality.

The extensive results and energy flow analysis establish reliability in the proposed method to serve as (close-to) optimal benchmark in real-world comprehensive and dynamic powertrain energy management problems, especially when globally-optimal Dynamic Programming will fail to remain computationally tractable.

\section{Conclusion}
In this paper, a comprehensive and large 13-state 4-control case-study problem for powertrain energy management of a class-6 parallel P2 hybrid electric truck
is solved using pseudo-spectral collocation based PS3 algorithm presented in our prequel \cite{paper1citation}.
A 20-minutes duty cycle for urban pickup and delivery application is used.
Complex nature of the real-world validated powertrain models
exhibiting discontinuous dynamics (engine on/off and gear selection), combinatorial constraints (minimum dwell-time), eco-driving control to modulate speed profile, thermal models of battery and after-treatment system, system-out fuel and NOx emission, various efficiency maps of engine, electric machine, and after-treatment emission models. Simulations are conducted for offline backward simulator with \textit{apiori} drive cycle information.

Detailed simulation results are presented for three problem cases: \textit{Fuel problem}, which only minimizes fuel consumption; \textit{Emissions problem}, which only minimizes system-out NOx emissions; and \textit{Fuel \& Emissions problem}, which has the two conflicting terms of fuel consumption and system-out NOx emissions weighed ideally based on a Pareto-front study. Trajectories of various dynamic signals are analyzed to capture the influence of every subsystem (transmission, engine, electric machine, after-treatment, battery, eco-driving controller) on the cumulative energy footprint | fuel and NOx emissions. Finally, comparative energy analysis is presented to establish the capability of serving as a benchmark optimal solution for the strong parallel powertrain problem.
In comparison to a coarsely modeled baseline solution, the Pareto-optimal result saves 7\% more fuel, reduces pollutant NOx emissions by 29\%, and demands 6\% lower energy from the powertrain system.


\bibliographystyle{IEEEtran}
\bibliography{IEEEabrv,biblibrary}






\end{document}